\documentclass[apj]{emulateapj}

\usepackage{color} 

\usepackage{lineno}

\def\ergs{erg~s$^{-1}$}
\def\ergcms{erg~cm$^{-2}$~s$^{-1}$}

\def\flam{erg~cm$^{-2}$~s$^{-1}$~\AA$^{-1}$}

\begin{document}
\title{Chandra and HST Observations of the Supersoft ULX in NGC 247: Candidate for Standard Disk Emission}

\author{Lian Tao\altaffilmark{1}, Hua Feng\altaffilmark{1}, Philip Kaaret\altaffilmark{2}, Fabien Gris\'e\altaffilmark{2}, Jing Jin\altaffilmark{1}}

\altaffiltext{1}{Department of Engineering Physics and Center for Astrophysics, Tsinghua University, Beijing 100084, China}
\altaffiltext{2}{Department of Physics and Astronomy, University of Iowa, Van Allen Hall, Iowa City, IA 52242, USA}

\shorttitle{Chandra and HST observations of NGC 247 ULX}
\shortauthors{Tao et al.}

\begin{abstract}
We report on multiwavelength observations of the supersoft ultraluminous X-ray source (ULX) in NGC 247 made with the Chandra X-ray Observatory and Hubble Space Telescope (HST). We aligned the X-ray and optical images using three objects present on both and identified a unique, point-like optical counterpart to the ULX. The X-ray to optical spectrum is well fitted with an irradiated disk model if the extinction measured for Cepheids in NGC 247 is used. Assuming only Galactic extinction, then the spectrum can be modeled as a standard thin accretion disk. Either result leads to the conclusion that a disk interpretation of the X-ray spectrum is valid, thus the source may be in the X-ray thermal state and contain an intermediate mass black hole of at least 600~$M_\sun$. In contrast to other supersoft ULXs which are transient and exhibit a luminosity temperature relation inconsistent with a disk interpretation of the X-ray emission, the NGC 247 ULX has a relatively steady flux and all available X-ray data are consistent with emission from a disk in the thermal state.

\end{abstract}

\keywords{accretion, accretion disks -- black hole physics -- galaxies: individual (NGC 247) -- X-ray: binaries}

\section{Introduction}

Ultraluminous X-ray sources (ULXs) are off-nuclear accreting black holes (BHs) with isotropic luminosities well above the classical Eddington limit of normal stellar mass BHs. Powering such high apparent luminosities requires accretion onto either ordinary stellar mass BHs ($M \sim10M_\sun$) with strong beaming and/or super-Eddington luminosities, massive stellar BHs ($M \la100M_\sun$) with mild beaming and/or near Eddington luminosities, or intermediate mass BHs (IMBHs; $10^2-10^4M _\sun$). An IMBH with quasi-isotropic sub-Eddington radiation is the preferred explanation for at least two ULXs \citep{fen10,ser11}. Thus, ULXs play a key role in the study of physics under extreme accretion, and comprise a promising population for the search for IMBHs which are important for the study of stellar evolution and the formation of supermassive black holes. Please see \citet{fen11} for a recent review.

Great interest has been focused on a subclass of ULXs that exhibit supersoft spectra and dramatic flux variability from $10^{37}$~\ergs\ up to nearly $10^{40}$~\ergs. Their X-ray spectra are dominated by photons below 2 keV, and the soft emission component is thermal-like with a temperature of around 0.1~keV. Due to their high peak luminosity and chaotic variability, they differ from canonical supersoft sources which could be nuclear-burning white dwarfs or X-ray supernova remnants \citep{van92, imm03}. There is no clear physical interpretation of supersoft ULXs. Either stellar mass BHs viewed at high inclination or IMBHs with cool disks (like in the thermal state) could account for some, but not all, of the observational facts.

Typical supersoft ULXs that have been described in detail include Antennae X-13 \citep{fab03}, M101 ULX-1 \citep{pen01,kon05,muk05}, M81 ULS1 \citep{swa02,liu08}, NGC 4631 X1 \citep{car07,sor09}, and NGC 247 ULX \citep{jin11}. The flux versus temperature relation of supersoft ULXs is the strongest argument against the IMBH scenario, as the supersoft ULXs show a nearly constant temperature while the flux varies by up to a factor of $10^3$.  This is strongly inconsistent with the observed properties of accretion flows around Galactic BH binaries. Specifically, it is inconsistent with the $L \propto T^{4}$ relation expected when the emitting surface area is fixed as expected for an accretion disk terminating at the innermost stable circular orbit. NGC 247 ULX is unusual amongst the supersoft ULXs: it has never been observed in a low flux state -- the observed X-ray flux has varied by no more than a factor of 3 in the 0.1--2 keV band in the available observations \citep{jin11}. Thus, NGC 247 ULX is a candidate IMBH and worth deep investigation, particularly on its multiwavelength emission, which has been found useful to constrain binary evolution and overall disk modeling.

The NGC 247 ULX showed an unabsorbed luminosity in the 0.3--10 keV band reaching up to $8 \times 10^{39}$~\ergs\ and clear variability in both short and long timescales \citep{win06,jin11}. The dominant component in its emission spectrum can be described by a blackbody spectrum with a temperature of about 0.1~keV \citep{rea97,win06,jin11}. In the 2009 XMM-Newton observation, which was so far the deepest one for the object, a weak but significant power-law component with a photon index around 2.5 was detected above 2~keV and an absorption feature around 1 keV was necessary to adequately fit the spectrum \citep{jin11}.

We obtained a Chandra observation to precisely determine the ULX X-ray position and HST imaging with a few filters to identify its optical counterpart and measure color and magnitude information. The observations and results are described in Section~\ref{sec:obs} and discussion of the physical interpretation is presented in Section~\ref{sec:dis}. A distance of 3.4~Mpc \citep{gie09} to NGC 247 is adopted in the paper.

\section{Observations, Analyses, and Results}
\label{sec:obs}

\subsection{Observations}

The Chandra observation (ObsID 12437; PI: H.\ Feng) was made on 2011 February 1 with an exposure of 5~ks. The focal plane instrument was the Advanced CCD Imaging Spectrometer (ACIS). About 5\arcmin\ to the north west of NGC 247 ULX lies a known background quasi-stellar object (QSO) PHL~6625, which was proposed as a reference object to align the Chandra and HST images, following the successful use of this technique to identify the optical counterpart of IC 342 X-1 \citep{fen08}. The Chandra telescope was pointed at the mid-point of the ULX and QSO and an offset of 1.1\arcmin\ was applied to the detector position so that the aimpoint fell toward the center of ACIS-S3 and both sources were imaged on the same chip.

The Advanced Camera for Surveys (ACS) with the Wide Field Channel (WFC) was adopted for the HST observations (Proposal ID 12375; PI: H.\ Feng) which were made on 2011 October 11 in two orbits. The field of view of ACS WFC is not large enough to cover both the ULX and QSO. Two exposures with an overlapped region were used to create a mosaic image. The F606W filter was selected because the archive contained images of nearby regions in this filter that could be used to expand the mosaic. Exposures with the F435W and F658N filters were also taken around the ULX. All observations are listed in Table~\ref{tab:obs}.

\begin{deluxetable}{ccccc}
\setlength{\tabcolsep}{\columnwidth}
\tabletypesize{\scriptsize}
\tablecaption{Chandra and HST observations of fields near NGC 247 ULX \label{tab:obs}}
\tablehead{
&  \colhead{ObsID}  & \colhead{Date} & \colhead{Instrument}  &\colhead{Exp.} \\
&  & & & \colhead{(s)}
}
\startdata
{\scriptsize Chandra}          & 12437 & 2011-02-01 & ACIS-S & 4988 \\
\noalign{\smallskip}\hline\noalign{\smallskip}
                 & j9ra77020 & 2006-09-20 & ACS/WFC/F606W & 1507   \\
                 & j9ra78020 & 2006-09-21 & ACS/WFC/F606W & 1507   \\
HST              & jblm01020$^{\star}$ & 2011-10-11 & ACS/WFC/F435W & 904    \\
                 & jblm01030$^{\star}$ & 2011-10-11 & ACS/WFC/F606W & 846    \\
                 & jblma1010           & 2011-10-11 & ACS/WFC/F606W & 846    \\
                 & jblm01010$^{\star}$ & 2011-10-11 & ACS/WFC/F658N & 1200
\enddata
\tablecomments{$^{\star}$that contains the ULX.}
\end{deluxetable}

\subsection{Data reduction}

With CIAO 4.4 and CALDB 4.4.7, a new level 2 events file was created for the Chandra observation using the {\tt chandra\_repro} script. Source detection was done with the {\tt wavdetect} tool on the 0.3--8 keV image. 172 net photons were detected from the ULX. Its X-ray spectral shape and flux, though not well constrained, are consistent with those measured with the 2009 XMM-Newton observation.

\begin{figure}[h!]
\centering
\includegraphics[width=\columnwidth]{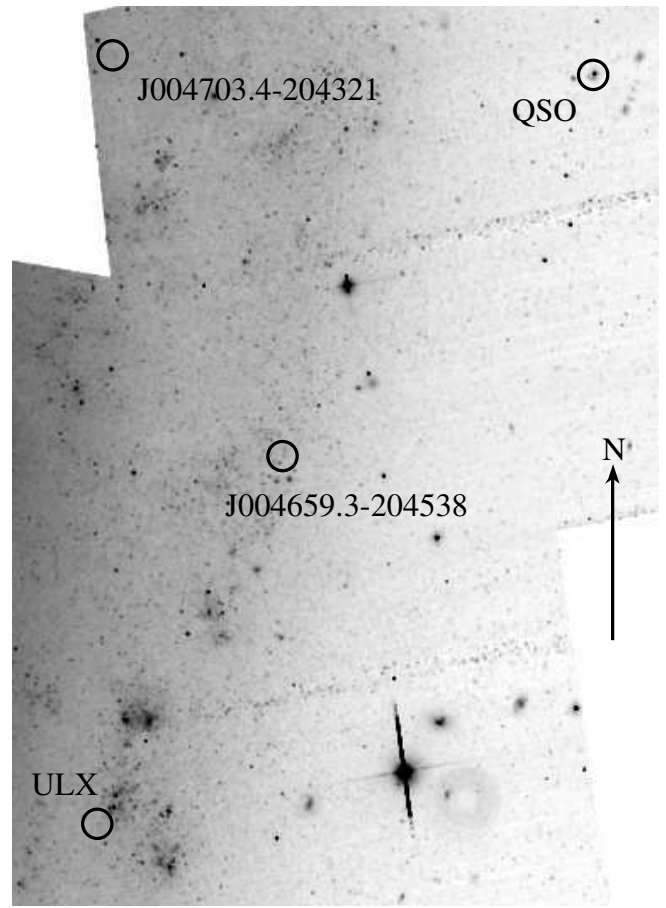}
\caption{The mosaic ACS F606W image of NGC 247 with 15-pixel Gaussian smoothing. The arrow points north and has a length of 1\arcmin.
\label{fig:mos}}
\end{figure}

For HST observations, the flat-fielded (\_flt) images delivered from the standard pipeline were adopted for data reduction. The cosmic ray flags in the data quality array were removed to avoid leaking into the final drizzled image if there is an overlap. Except for the two observations in 2006 from the archive, the bias striping noise was removed using {\tt acs\_destripe} and correction to the charge transfer efficiency was done using {\tt PixCteCorr}. Then, drizzled images were created using the {\tt multidrizzle} task, in which cosmic ray removal was executed. For the four F606W images, their relative transformations were calculated with the {\tt geomap} task using objects in the overlapped region found by {\tt daofind} and {\tt xyxymatch}. The {\tt multidrizzle} task was used to create the mosaic image that is shown in Figure~\ref{fig:mos}, which was aligned to the Two Micron All Sky Survey \citep[2MASS;][]{skr06} grid for better absolute astrometry.

\subsection{Astrometry and optical identification}

\begin{figure*}
\centering
\includegraphics[width=0.8\textwidth]{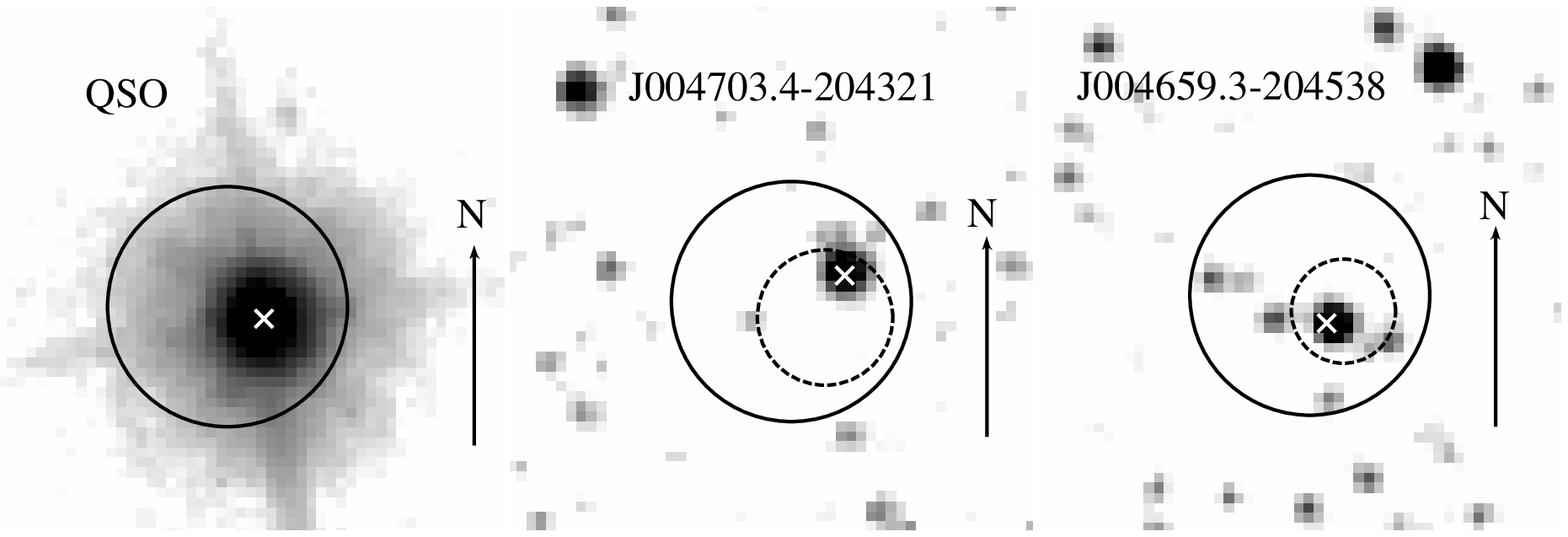}\\
\includegraphics[width=0.8\textwidth]{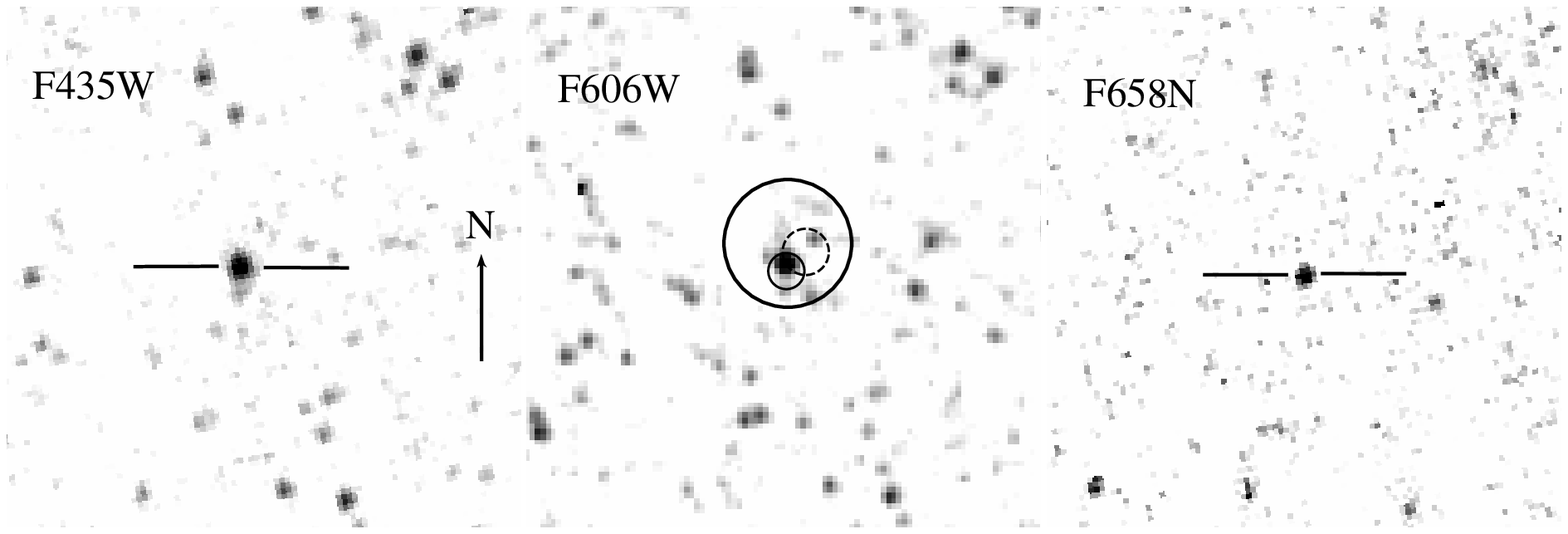}
\caption{HST images around the reference objects (upper panels) and the ULX (lower panels). Each large solid circle indicates the original Chandra position for the indicated source with an error radius of 0.6\arcsec. The dashed circles indicate the X-ray positions corrected in translation only by alignment with the QSO. The crosses indicate the X-ray positions of the reference objects after correction for translation, scale, and rotation. The smaller solid circle on the F606W image indicates the corrected X-ray position after alignment using the three reference objects. The horizontal bars on the F435W and F658N images indicate the identified counterpart of the ULX. North is up and the length of the arrow is 1\arcsec.
\label{fig:opt}}
\end{figure*}


\begin{deluxetable}{cllll}
\tablecolumns{6}
\tablewidth{0pc}
\tabletypesize{\scriptsize}
\tablecaption{X-ray and optical positions of the reference objects and the ULX \label{tab:cpt}}
\tablehead{
\colhead{Source}&\colhead{}&\colhead{R.A.} &\colhead{decl.}        &\colhead{err} \\
\colhead{}      &                &\colhead{(J2000.0)}&\colhead{(J2000.0)}&\colhead{(\arcsec)}
}
\startdata
QSO              & X   & 00 46 51.823 & $-$20 43 28.30 & 0.104 \\
                 & X$^\prime$ & 00 46 51.810 & $-$20 43 28.36 &       \\
                 & O      & 00 46 51.811 & $-$20 43 28.38 & \\

\hline\noalign{\smallskip}

J004659.3$-$204538 & X    & 00 46 59.376 & $-$20 45 38.49 &0.219 \\
                 & X$^\prime$  & 00 46 59.370 & $-$20 45 38.63 &  \\
                 & O       & 00 46 59.368 & $-$20 45 38.63 &       \\
\hline\noalign{\smallskip}
J004703.4$-$204321 & X    & 00 47 03.508 & $-$20 43 21.63 &0.308 \\
                 & X$^\prime$  & 00 47 03.489 & $-$20 43 21.50 &      \\
                 & O       & 00 47 03.490 & $-$20 43 21.48 &      \\
\hline\noalign{\smallskip}
ULX              & X    & 00 47 03.882 & $-$20 47 44.04 &0.074 \\
                 & X$^\prime$  & 00 47 03.883 & $-$20 47 44.30 &0.17  \\
                 & O       & 00 47 03.884 & $-$20 47 44.23 &
\enddata
\tablecomments{
X -- original Chandra position derived from {\tt wavdetect};
X$^\prime$ -- corrected X-ray position by alignment with the three reference objects;
O -- position of the optical counterpart on the mosaic F606W image. The error radius is quoted at 90\% confidence, converted from 68\% intervals at 1-dimension assuming Rayleigh distribution.
}
\end{deluxetable}

Alignment of the Chandra and HST images was performed in two steps. First, the background QSO was used to register the two images. The relative shift can be corrected in this way but scale and rotation errors remain. Besides the ULX and QSO, there were 5 X-ray sources detected in the field of the mosaic image. Two of them with a significance over 5$\sigma$ have a unique optical source within 0.6\arcsec, which are XMMU~J004659.3$-$204538 and XMMU~J004703.4$-$204321 \citep{jin11}. Taking into account of the scale and rotation uncertainties of Chandra images\footnote{http://cxc.harvard.edu/cal/Hrma/PlateScale.html}, which are typically 0.0001\arcsec/pixel and $0.01^{\circ}$, respectively, at the 68\% confidence level, we identified unique optical counterparts.  These are shown in Figure~\ref{fig:opt}, where the solid circles indicate the Chandra original positions with an absolute uncertainty of 0.6\arcsec\ and the dashed circles represent the corrected positions with error circles after alignment using the QSO.

Then, the QSO, J004659.3$-$204538, and J004703.4--204321 were employed as reference objects to further align the images, which allowed us to fit the relative shift, scale, and rotation between the two images.  Using the {\tt geomap} tool, fitting the transformation resulted in a root-mean-square of 0.02\arcsec, indicative of adequate fitting. The new relative position error for the ULX is mainly from statistical errors on the X-ray positions of the reference sources, which were added quadratically. The systematic error due to the asymmetry of the point spread function (PSF) is insignificant according to simulation with MARX. The relative position error of the ULX after alignment is 0.17\arcsec, leading to a unique identification in optical. All positions and uncertainties are listed in Table~\ref{tab:cpt}.

\subsection{HST Photometry}
\label{sec:phot}


\begin{deluxetable}{cccc}
\tablewidth{0pc}
\tabletypesize{\scriptsize}
\tablecaption{Fluxes and magnitudes of NGC 247 ULX \label{tab:flux}}
\tablehead{
\colhead{Band} & \colhead{$F_{\lambda} / 10^{-18}$}  &\colhead{$m_0$} & \colhead{$m_0$} \\
&\colhead{} & \colhead{$E(\bv) = 0.018$} & \colhead{$E(\bv) = 0.18$}
}
\startdata
F435W &  $5.49\pm0.17$ & $22.60\pm0.03$  & $21.92\pm0.03$ \\
F606W &  $2.68\pm0.13$ & $22.52\pm0.05$  & $22.04\pm0.05$ \\
F658N &  $2.1\pm0.2$ & \nodata           & \nodata        \\
$B$   &  \nodata     &  $22.61\pm0.03$   & $21.93\pm0.03$ \\
$V$   &  \nodata     &  $22.58\pm0.07$   & $22.06\pm0.07$ \\
$\bv$ &  \nodata     &  $0.03\pm0.07$    & $-0.13\pm0.07$ \\
$M_V$ &  \nodata     & $-5.08\pm0.13$    & $-5.60\pm0.13$ \\

$\alpha$ & \nodata & $0.34\pm0.18$ & $0.94\pm0.18$
\enddata
\tablecomments{$F_\lambda$ is the observed flux in units of \flam; The magnitudes have Vega zeropoints and are extinction corrected assuming different values. $\alpha$ is the two-point power-law index ($F_\nu \propto \nu^\alpha$) between F435W and F606W.}
\end{deluxetable}

For the ULX, aperture photometry was performed on each image using the IRAF package APPHOT. The {\tt calcphot} task in the SYNPHOT package was used for aperture correction and flux conversion.  The Galactic extinction along the direction of NGC 247, estimated from the COBE dust map \citep{sch98}, is $E(\bv)=0.018$ mag, which is a lower limit. \citet{gie09} reported a mean extinction of $E(\bv) = 0.18$ in NGC 247 via the optical Cepheids studies, indicative of significant extinction within the host galaxy. The $E(\bv)$ values derived from two Cepheids (cep008 and cep018) near the ULX region are 0.17 and 0.22, respectively, close to the mean value. The X-ray neutral hydrogen column density $N_{\rm H}$ of the ULX was found to be around $3.0 \times 10^{21}$ cm$^{-2}$ \citep{jin11}. Following the correlation between $N_{\rm H}$ and extinction \citep{pre95}, we obtained $E(\bv) = 0.54$ assuming $R_V = 3.1$. Usually, the extinction derived from X-ray absorption is an overestimate and should be regarded as an upper limit. For the ULX, an extinction of $E(\bv) = 0.54$ is too high to be physical as it leads to an optical spectrum steeper than the Rayleigh-Jeans law ($F_\nu \propto \nu^2$). Thus, we only consider the extinctions $E(\bv) = 0.018$ and 0.18 in the following. The observed flux density, extinction corrected magnitudes in the observed bands and the standard Johnson $B$ and $V$ bands, which moderately match the F435W and F606W throughputs, respectively, are listed in Table~\ref{tab:flux}.

In the F658N filter, which is a narrow band including the H$\alpha$ and [N~{\sc ii}] lines, the observed flux density is $(2.1 \pm 0.2) \times 10^{-18}$~\flam. Flux from the F606W filter, with a band that contains the F658N band, was used to estimate the continuum contribution in the narrow band. Assuming the continuum has a power-law spectrum fitting the F435W and F606W fluxes, we estimate that the continuum contributed $(2.1 \pm 0.2) \times 10^{-18}$~\flam\ in the F658N band.  This suggests that no H$\alpha$ line flux is detected.  We note there is no nebulosity around the ULX apparent in the F658N image.

\begin{figure*}
\centering
\includegraphics[width=0.8\textwidth]{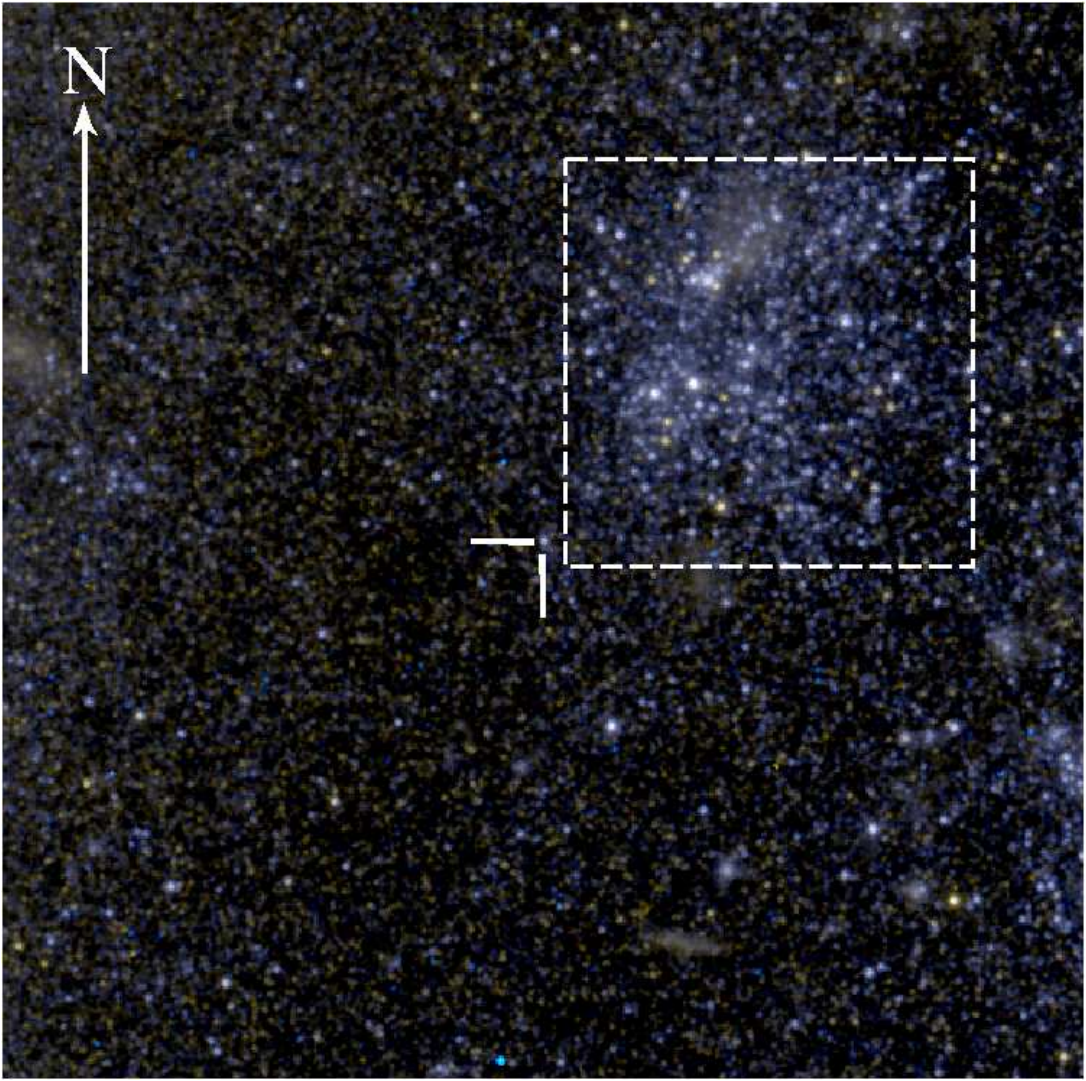}
\caption{HST ACS WFC color composite around NGC 247 ULX (red = F606W, green = (F606W+F435W)/2, and blue = F435W) with a size of $40\arcsec \times 40\arcsec$. The box with a size of $15\arcsec \times 15\arcsec$ indicates the region of a nearby stellar association. The bars indicate the ULX optical counterpart. The arrow has a length of 10\arcsec and points north.
\label{fig:color}}
\end{figure*}

\begin{figure}[thb]
\centering
\includegraphics[width=\columnwidth]{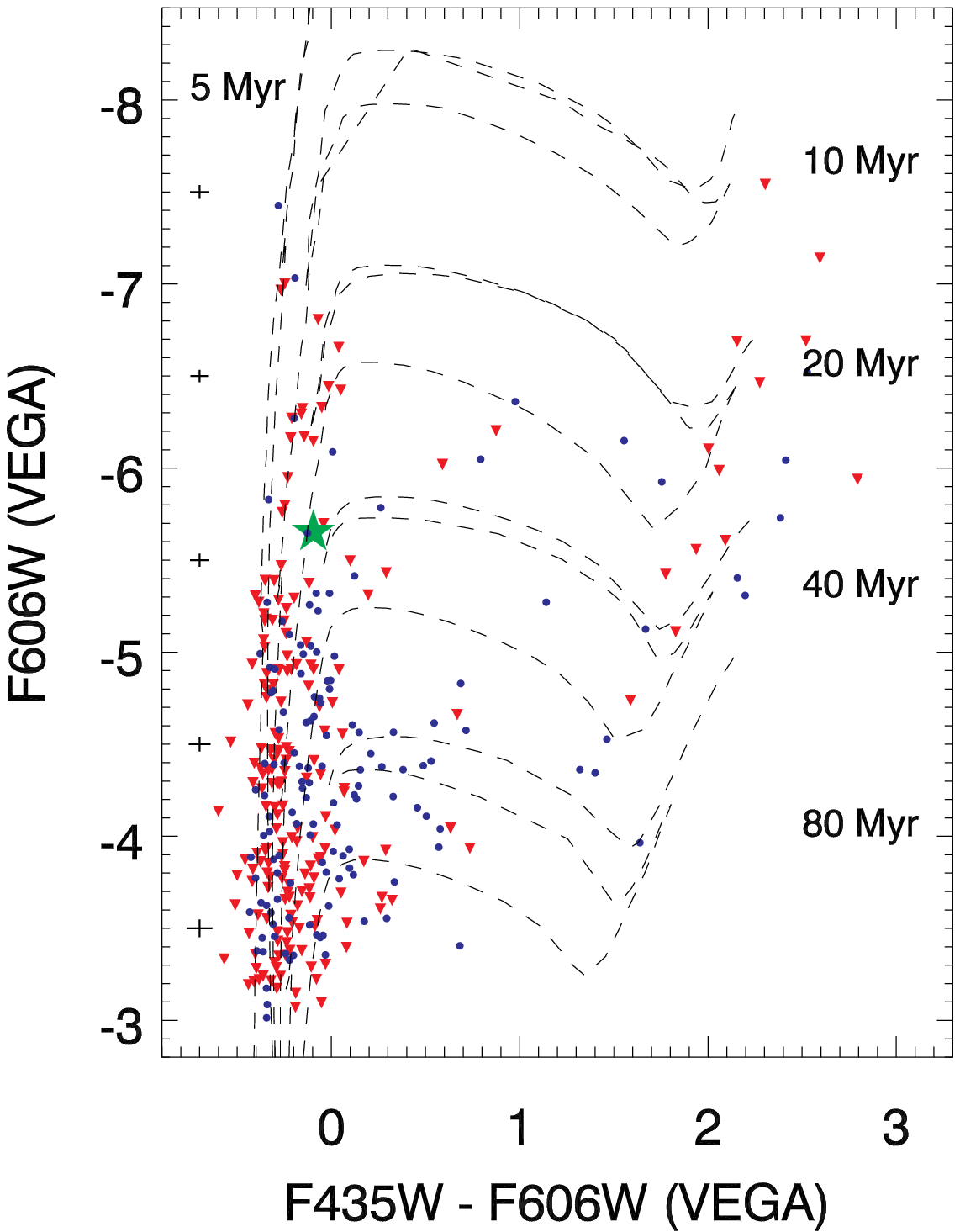}
\caption{Color magnitude diagram for stars within a 20\arcsec-radius circular field around NGC 247 ULX. The magnitudes have Vega zeropoints corrected for an extinction of $E(\bv)=0.18$. Typical errors at various magnitudes are shown on the left. Red triangles indicate sources in the stellar association and blue dots indicate sources in the field. The green star indicates the ULX counterpart. Isochrones of 5, 10, 20, 40, and 80 Myr are shown.
\label{fig:cmd}}
\end{figure}

Figure~\ref{fig:color} shows a color composite image around the region of NGC 247 ULX, produced using the F435W and F606W images. The ULX appears near a loose stellar association with a size of $\sim 15\arcsec \times 15\arcsec$ ($\sim 250 {\rm pc} \times 250 {\rm pc}$). We performed PSF photometry for stars within a $20\arcsec$-radius stellar field around NGC 247 ULX. Objects with flux below 10 times sky deviation or with $\chi^2 > 3$ during PSF fits are excluded. Then, applying Vega zero points, the dereddened magnitude are calculated assuming $E(\bv) = 0.18$ and plotted in the the color magnitude diagram (CMD), see Figure~\ref{fig:cmd}. Typical errors are 0.04 for the magnitudes and 0.06 for the colors in the range $-8 \lesssim M_{\rm F606W} \lesssim -3$. No spectroscopic metallicity has been reported for NGC 247. Following \citet{dav06}, we adopt $Z=0.008$ based on the integrated brightness of the galaxy and CMD analysis. The isochrones\footnote{http://stev.oapd.inaf.it/cgi-bin/cmd} of 5, 10, 20, 40, and 80 Myr \citep{mar08,gir10} are also shown in Figure~\ref{fig:cmd}. Stars in either the association or in the field show a continual star formation history to the present date, but most bright stars in the association have ages consistent with 10--20 Myr. The field stars are, in general, slightly dimmer and older than those in the association.

\subsection{Multiwavelength modeling}

\begin{deluxetable*}{cllllcllcccl}
\tablewidth{\textwidth}
\tablecolumns{12}
\setlength{\tabcolsep}{3pt}
\tabletypesize{\scriptsize}
\tablecaption{Spectral fitting of NGC 247 ULX \label{tab:xpara}}
\tablehead{
\colhead{Model} &\colhead{$N_{\rm H}$} &\colhead{$E$} &\colhead{$\tau$} &\colhead{$kT$} & \colhead{$R_{\rm in} \rm \sqrt{cos~\theta}$} & \colhead{$\Gamma$} & \colhead{$N_{\rm PL}$}&\colhead{$f_{\rm out}$} & \colhead{$\log(r_{\rm out}/r_{\rm in})$} & \colhead{$L_{\rm c}/L_{\rm d}$} &\colhead{$\chi^2/{\rm dof}$}
}
\startdata
1 & $3.2^{+0.7}_{-0.7}$ & $1.02^{+0.02}_{-0.03}$ & $1.4^{+0.5}_{-0.5}$ & $0.13^{+0.02}_{-0.02}$   & $0.9^{+1.3}_{-0.5}$ & $2.5^{+1.4}_{-1.0}$ & $1.7^{+3.8}_{-1.2}$ &\nodata & \nodata & \nodata & 36.8/40 \\
2 & $2.7^{+1.0}_{-0.7}$ & $1.02^{+0.03}_{-0.03}$ & $1.3^{+0.5}_{-0.4}$ & $0.112^{+0.016}_{-0.017}$& $1.0^{+1.3}_{-0.5}$ & $2.7^{+2.2}_{-1.0}$ & $2.0^{+7.5}_{-1.4}$ &\nodata & \nodata & \nodata & 39.0/40 \\
3 & $3.2^{+1.0}_{-0.8}$ & $1.02^{+0.03}_{-0.02}$ & $1.4^{+0.5}_{-0.4}$ & $0.13^{+0.02}_{-0.02}$   & $1.0^{+1.4}_{-0.5}$ & $2.5^{+1.4}_{-1.0}$ & \nodata             &$0.015^{+0.039}_{-0.014}$ & $3.2 \pm 0.4$ & $0.017^{+0.076}_{-0.006}$ & 36.6/40
\enddata
\tablecomments{
Model 1: wabs $\ast$ edge $\ast$ (diskbb + powerlaw);
Model 2: wabs $\ast$ edge $\ast$ (bbodyrad + powerlaw);
Model 3: redden $\ast$ wabs $\ast$ edge $\ast$ diskir.
$N_{\rm H}$ is the X-ray absorption column density in $10^{21}~\rm cm^{-2}$;
$E$ is the absorption edge in keV;
$\tau$ is the optical depth;
$kT$ is the temperature of the thermal component in keV;
$R_{\rm in}$ is the inner radius in $10^{4}$ km;
$\theta$ is the inclination angle;
$\Gamma$ is the power-law photon index;
$N_{\rm PL}$ is the power-law normalization at 1 keV in $10^{-5}$ photons keV$^{-1}$cm$^{-2}$s$^{-1}$;
$f_{\rm out}$ is the fraction of bolometric luminosity thermalized in the outer disk;
$r_{\rm out}$ and $r_{\rm in}$ are respectively the outer and inner disk radius;
$L_{\rm c}/L_{\rm d}$ is the ratio of flux between the Compton component and the un-illuminated disk.
All errors and limits are at 90\% confidence level.
Model 1 \& 2 are quoted from \citet{jin11}.
}
\end{deluxetable*}

No simultaneous X-ray and optical observations are available for the ULX. In order to evaluate the X-ray to optical flux ratio, we estimated the X-ray flux range from two XMM-Newton observations \citep{jin11}. $\log(f_{\rm X}/f_V)$ is defined as $\log f_{\rm X} + m_V/2.5 + 5.37$, where $f_{\rm X}$ is the observed X-ray flux in the 0.3-3.5 keV band in $\rm ergs~cm^{-2}~s^{-1}$ and $m_V$ is the observed visual magnitude \citep{mac82}. We adopted $f_{\rm X} = (1.6 - 5.8) \times 10^{-13}$~\ergcms\ and $m_V = 22.64$ and obtained $\log(f_{\rm X}/f_V) = 1.6 - 2.2$, which is larger than that of active galactic nuclei, clusters of galaxies, normal galaxies and normal stars, and is marginally consistent with the value for BL Lac objects. A blazar nature can ruled out for this source due to its supersoft spectrum. Therefore, NGC 247 ULX is most likely an X-ray binary \citep{mac88,sto91}.

\begin{figure*}
\centering
\includegraphics[width=\textwidth]{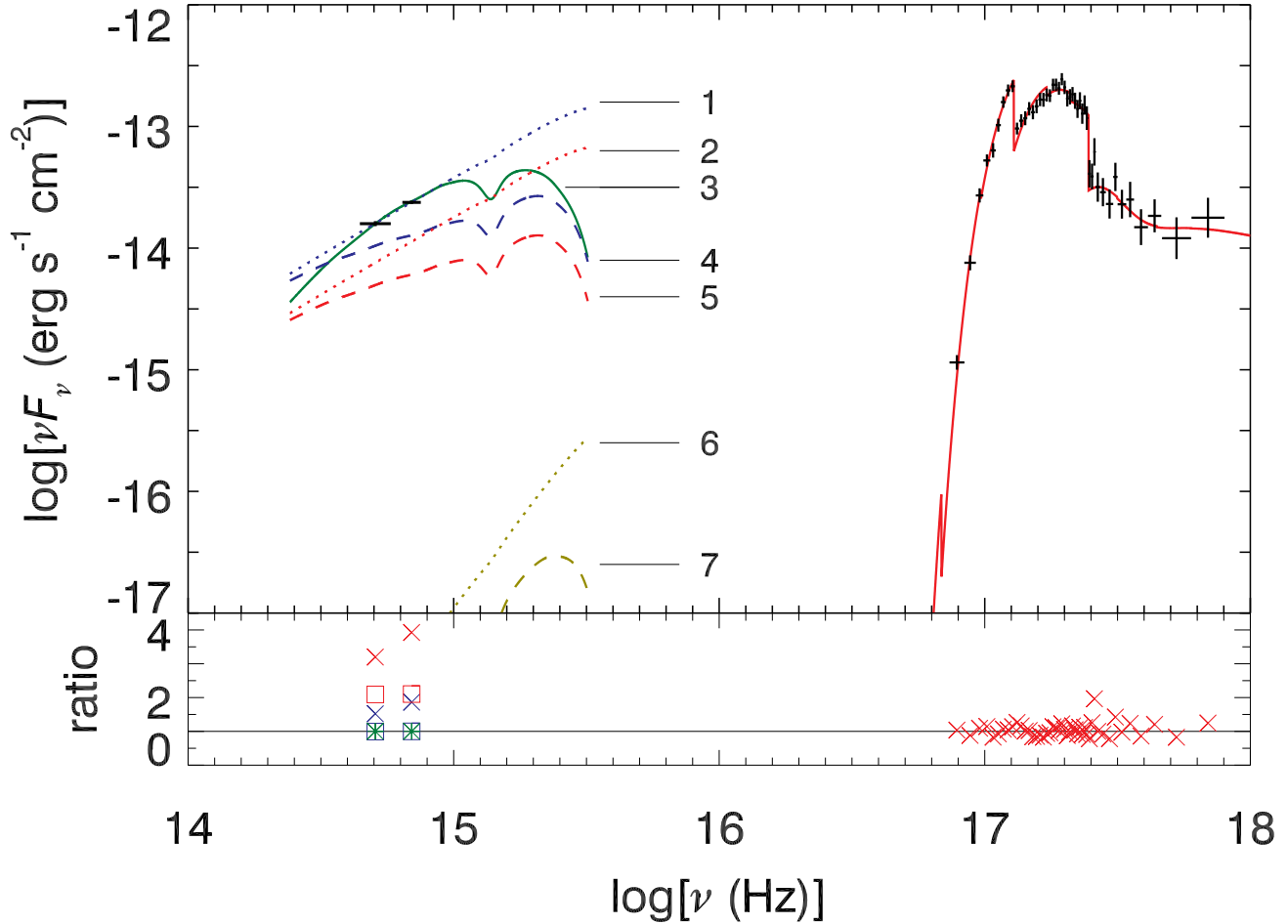}
\caption{Observed broad band spectrum of NGC 247 ULX. The X-ray data are extracted from the 2009 XMM-Newton observation and the red solid curve is the best-fit model, see \citet{jin11} for details. In the optical band, the two black points are HST observations. Numbered lines correspond to:
(1) extrapolation of the disk blackbody model for the 2001 XMM-Newton observation assuming $E(\bv) = 0.018$;
(2) same as (1) but for the 2009 observation.
(3) best-fit model with {\tt diskir} assuming $E(\bv) = 0.18$;
(4) same as (1) but assuming $E(\bv) = 0.18$;
(5) same as (2) but assuming $E(\bv) = 0.18$;
(6) and (7) are extrapolations of the mono-temperature blackbody model assuming the two extinctions.
\label{fig:spec}}
\end{figure*}

The X-ray spectrum of the ULX can be decomposed into two components, a dominant soft thermal component (either blackbody or disk blackbody) with a temperature about 0.1~keV plus a weak hard power-law component \citep{jin11}, see Table~\ref{tab:xpara}. The mono-temperature blackbody component underestimates the optical flux by 4 orders of magnitude, as shown in Figure~\ref{fig:spec}. In contrast, the optical flux extrapolated from the disk blackbody model is consistent with or close to the observations, depending on the assumed extinction. For the 2001 XMM-Newton observation, the predicted optical flux is fully consistent with that observed assuming $E(\bv) = 0.018$ and is 1.5--2 times lower than observed assuming $E(\bv) = 0.18$.  The discrepancy is 2 times larger for the 2009 XMM-Newton observation. The two-point power-law index between the F435W and F606W bands is $0.34 \pm 0.18$ assuming $E(\bv) = 0.018$ and $0.94 \pm 0.18$ assuming $E(\bv) = 0.18$, see Table~\ref{tab:flux}. For the standard accretion disk model, the apparent power-law index in the intermediate band (i.e.\ the optical band for this case) is 1/3. Thus, with the assumption of Galactic extinction, both the optical flux and spectral shape are consistent with that predicted for the disk model. For an extinction of $E(\bv) = 0.18$, the inconsistent flux could be explained by variability, however, the spectral index would be significantly higher than 1/3 and difficult to explain with a standard disk model.

Assuming $E(\bv) = 0.18$, we attempted to fit the X-ray and optical data with a Comptonized, irradiated disk model \citep[{\tt diskir};][]{gier09}, which has been applied successfully to several ULXs, e.g.\ NGC 5408 X-1, Holmberg II X-1 and HLX-1 \citep{kaa09, gri12, tao12, sor12}. The 2009 XMM-Newton observation was adopted, which was the only one with sufficient statistics for detection of the Compton tail \citep{jin11}, and the spectrum was extracted and grouped following \citet{jin11}. Besides a soft thermal component and a Compoton tail, an absorption feature near 1~keV, close to the highly ionized Fe-L edge, was added to achieve an adequate fit. The feature may be due to highly ionized accretion disk winds that are found ubiquitously in the soft state of Galactic BHs \citep{pon12}. The XMM-Newton spectrum and the two optical data points were fitted to a {\tt diskir} plus edge model subject to interstellar absorption using the chi square statistics. The multiband fitting parameters are listed in Table~\ref{tab:xpara} (Model 3) along with X-ray fitting parameters (Model 1 \& 2) quoted from \citet{jin11}. The parameters of interest are those affecting the optical emission, including the fraction of bolometric flux thermalized in the outer disk ($f_{\rm out}$) and the ratio of outer to inner disk radii ($r_{\rm out}/r_{\rm in}$). Scaling the X-ray luminosity to the normalization derived from the 2001 XMM-Newton observation, $f_{\rm out}$ becomes $0.003 \pm 0.001$ by fixing $\log(r_{\rm out}/r_{\rm in})$ at 3.2 (otherwise the parameters are poorly constrained). We conclude that for an extinction of $E(\bv) = 0.18$, as suggested by observations of Cepheids in the host galaxy, X-ray irradiation in the outer disk is able to produce both the high optical flux and steep optical spectral slope.

\section{Discussion}
\label{sec:dis}

The Chandra and HST images were properly aligned using three objects present on both, leading to a relative position error slightly less than 0.2\arcsec. This enables a unique identification of the optical counterpart of NGC 247 ULX as a point-like source lying near a loose stellar association. While most stars in the association brighter than the ULX counterpart have ages of 10--20~Myr (Figure~\ref{fig:cmd}), a few of them are consistent with younger (5~Myr) or older (40 Myr) populations. This suggests that the association may have undergone continual star formation and be  composed of OB associations formed at different epochs.

The most likely extinction to stars in the nearby stellar field, as described in Section~\ref{sec:phot}, is close to the value derived from Cepheids in the host galaxy, $E(\bv) = 0.18$. Because NGC 247 is a nearly face-on galaxy, we have no constraint on the location of the ULX with respect to the disk plane. In the Milky Way, most HMXBs are found in the Galactic plane while LMXBs tend to be widely distributed in the halo. If the source is in front of the disk plane, a Galactic extinction of $E(\bv) = 0.018$ is likely. The X-ray extinction ($E(\bv) = 0.54$) is ruled out by the physical constraint that no thermal spectrum can be steeper than the Rayleigh-Jeans law.

The source appears like a B8--A5 Ib supergiant assuming $E(\bv) = 0.018$, or a B1--B7 Ib assuming $E(\bv) = 0.18$. However, the apparent colors and magnitudes should not be used for classification of the companion star, as the optical light from ULXs could be dominated by emission on the accretion disk \citep{kaa05,tao11,gri11}.

A single temperature blackbody model fits the X-ray data well, and could be interpreted as emission from the photosphere of a massive outflow associated with super-Eddington accretion that shadows the inner disk and also the hard X-rays if viewed at high inclination angle. However, this model underestimates the optical flux by 4 orders of magnitude. Thus, in this scenario, the optical and X-ray emission must be due to different mechanisms. A disk with the inner part obscured or truncated, or the companion star, could be the origin of the optical emission. Future spectroscopic observations may be able to distinguish these two possibilities.

A disk blackbody model, in contrast, reasonably accounts for both the X-ray and optical observations. If one considers only the Galactic extinction, which implies that the source is nearer to us than the disk plane of NGC 247, then both the observed fluxes and spectral index in the optical band are fully consistent with the standard disk model extrapolated from the 2001 XMM-Newton observation. These suggest that we may have seen a pure (or dominant) standard thin accretion disk that extends from optical to X-rays, across a radial range of 3 orders of magnitudes, with little if any irradiation.

If this is the nature of the optical emission, it suggests the presence of an IMBH, and the outer disk radius needs be at least 300 times the inner radius in order to have an optical spectrum unbroken at the F606W band. Based on the disk blackbody model (Table~\ref{tab:xpara} Model 1), the disk inner radius is derived to be $0.9^{+1.3}_{-0.5}\times 10^{4}$ km assuming a face-on disk, corresponding to a BH mass of $1200^{+1800}_{-600}~M_\sun$ using the Schwarzschild metric. This suggests that the ULX was accreting at 4\% of the Eddington limit during the 2009 XMM-Newton observation, or 10\% during the 2001 XMM-Newton observation, typical of accretion rates for Galactic black holes in the thermal state \citep{gier04}. Future optical and infrared observations are important to test such a scenario. In particular, simultaneous X-ray and optical observations would provide a strong test of the model and photometry or spectroscopy in the near-infrared band would be useful to determine the size of the accretion disk.

Adopting the higher extinction as measured for Cepheids in the host galaxy, the observed optical flux is a few times higher than extrapolation of the disk model, which can be regarded as being consistent with the degree of X-ray variability, since the Einstein, ROSAT, XMM-Newton, and Chandra observations show an apparent flux change by a factor of 3 in the 0.1-2 keV range. However, the two-point spectral index is no longer consistent with that predicted for the standard thin disk. Thus, disk irradiation is needed to account for the excess optical emission.

It is interesting to compare the results obtained for the irradiated disk model with those derived for NGC 5408 X-1, which is the ULX with the most detailed simultaneous X-ray and optical observations available \citep{gri12}. The disk contribution is one of the highest among ULXs and its optical flux is about 3--8 times higher than predicted for the disk model in the $V$ band. The luminosity fraction of the Comptonization component in NGC 5408 X-1 ranges around 0.4--0.7, consistent with the identification of the steep power-law state. Despite the large errors, the excess optical flux above the intrinsic disk emission, proportional to $f_{\rm out}$, was found to scale with the luminosity fraction of the Compton tail in the X-ray band \citep[Table~4 in][]{gri12}, similar to the results found for the Galactic BH binary XTE~J1817$-$330 while in the same state \citep[Figure~8 in][]{gier09}. For NGC 247 ULX, its Compton tail fraction is less than 0.1, consistent with in the thermal state. The inferred fractional luminosity thermalized in the outer disk, if the 2001 XMM-Newton observation is adopted, is consistent with values found in the thermal state of XTE~J1817$-$330 \citep{gier09}. In sum, both the X-ray and multiband modeling suggests that NGC 5408 X-1 is likely in the steep power-law state while NGC 247 ULX is in the thermal state.

Thus, we conclude that NGC 247 ULX shows no disk irradiation if there is only Galactic extinction, and some irradiation if there is extinction within the host galaxy. In both cases, a disk interpretation of the X-ray data and identification of the source as being in the X-ray thermal state are valid.

\begin{deluxetable}{lccclc}
\tablewidth{\columnwidth}
\tabletypesize{\scriptsize}
\tablecaption{Transient properties of supersoft ULXs \label{tab:tran}}
\tablehead{
 & \colhead{Obs.} & \colhead{\parbox[c]{5ex}{high state\vspace{0.1cm}}} & \colhead{\parbox[c]{5ex}{low state\vspace{0.1cm}}} & \colhead{$L_{\rm max}/L_{\rm min}$} & \colhead{Refs.}
}
\startdata
M101 ULX-1    &  33 & 11  & 22 & $>10^2$ & a \\
Antennae X-13 &  4  & 2   & 2  & $\sim 10^2$ & b \\
M81 ULS1      &  17 & 10  & 7  & $\sim 50$ & c  \\
NGC 4631 X1   &  8  & 5   & 3  & $\sim 10^3$ & d  \\
NGC 247 ULX   &  8  & 8   & 0  & $\sim 3$ & e
\enddata
\tablenotetext{a}{\citet{liu09}}
\tablenotetext{b}{\citet{fab03}}
\tablenotetext{c}{\citet{liu08}}
\tablenotetext{d}{\citet{vog96,car07,sor09}}
\tablenotetext{e}{\citet{fab92,jin11}}
\end{deluxetable}

As mentioned in the introduction, the strong argument against a thermal disk interpretation of supersoft ULXs is their transient behavior. In Galactic black hole binaries, the disk temperature, geometry, density, optical depth, and the emergent spectral shape will vary dramatically along with a flux change up to a factor of $10^3$. However, the measured temperature is similar for supersoft ULXs in their low and high states. In other words, they appear to be supersoft no matter how the flux changes. In contrast, NGC 247 ULX is never found in the low state. It has been observed 8 times: with Einstein (1), ROSAT (4), XMM-Newton (2), and Chandra (1). The measured X-ray flux in the 0.1--2 keV band varied only over a factor of 3 in these observations. In Table~\ref{tab:tran}, we list the number of low and high states and the ratio of maximum to minimum luminosity for 5 supersoft ULXs that have been extensively studied. The typical luminosity is $10^{39}$~\ergs\ in the high state and $10^{37}$~\ergs\ in the low state. The lack of low state in NGC 247 may suggest that it is unlike other supersoft ULXs, and the disk interpretation may still be valid. The only evidence that is potentially inconsistent with a disk interpretation for NGC 247 ULX is that it exhibits strong short-term variability \citep{jin11}, although the power spectrum density has the same shape with that seen in the thermal state of BH binaries. It is unkown whether or not the soft state at such a high luminosity has similar timing properties as in stellar mass BH binaries, which usually are at least 10 times dimmer.

Following \citet{gier09}, the effective temperature of the outer irradiated disk can be expressed as
\begin{displaymath}
T_{\rm eff} = T_{\rm in} \left\{ 
  r^{-3} + f_{\rm out}r^{-2} \left[1 + {L_{\rm c} \over L_{\rm d}} \left(1 + f_{\rm in}\right) \right]
 \right\} ^{1 \over 4}
\end{displaymath}
where $r=r_{\rm out}/r_{\rm in}$ and $f_{\rm in}$ is the fraction of Compton emission thermalized in the inner disk, which was fixed to be 0.1 in our fits. Using our best-fit spectral parameters, the outer disk temperature is estimated to be about 13000~K for NGC 247 ULX. This is higher than the Hydrogen ionization temperature of about 6500~K and suggests that the disk is not subject to disk instability \citep{kin96}. This may explain the fact that the ULX has always been observed in the high state for three decades. The derived outer disk radius of $\sim 10^{12}$~cm indicates that the Roche-lobe radius must be at least this large.

To summarize, the HST observations reveal that NGC 247 ULX is a good candidate to be an IMBH with dominant standard disk emission. Future observations from near-UV to near-infrared are needed to test the optical disk spectrum, and repeated X-ray observations of moderate depth are needed to test the $L_{\rm disk} \propto T_{\rm in}^4$ relation predicted for the thermal state. 

\acknowledgments We are grateful to the referee for useful comments that improved the paper. HF acknowledges funding support from the National Natural Science Foundation of China under grant No.\ 10903004 and 10978001, the 973 Program of China under grant 2009CB824800, the Foundation for the Author of National Excellent Doctoral Dissertation of China under grant 200935, the Tsinghua University Initiative Scientific Research Program, and the Program for New Century Excellent Talents in University.  PK and FG acknowledge partial support from STScI grant HST-GO-12375.01-A and Chandra grant GO1-12048X.

{\it Facility:} \facility{Chandra}, \facility{HST}, \facility{XMM-Newton}

\end{document}